\documentclass[twocolumn,pre]{revtex4}
\usepackage{amsmath,amsfonts,amssymb}

\usepackage{graphicx}           

\begin{document}

\preprint{LA-UR-XX-XXXX}

\title{Free Energy Perturbation Theory at Low Temperature}

\author{C. W. Greeff}

\affiliation{Los Alamos National Laboratory, Los Alamos, NM 87545}

\date{\today}

\begin{abstract} The perturbative expansion introduced by 
Zwanzig [R. W. Zwanzig, J. Chem. Phys. {\bf 22}, 1420 (1954)]
expresses the difference in Helmholtz free energy between a system of 
interest and that of a reference system as series of cumulants $\kappa_n$
of the potential energy difference between the two systems. 
This expansion has attractive features as a method for obtaining 
absolute free energies for {\it ab initio} potential energy surfaces.
The series is formally a power series in $\beta=1/T$, suggesting
that its usefulness may be limited to high temperature.
However, for smooth reference potentials, the $T$-dependence of 
the $\kappa_n$ contributes to the convergence.
A closed form expression is derived for the $\kappa_n$ to all orders
for the case that both the system and reference potentials are harmonic.
In this case $\kappa_n \propto T^n$ for $n \ge 2$ and the convergence of
the series is independent of temperature. More realistic cases of liquid
Cu and solid Al, with a $1/r^{12}$ and harmonic reference potential, 
respectively, are investigated numerically by evaluating the cumulants 
to third order using Monte Carlo integration. In all cases, the ratio
of the third order to the second order term in the expansion is found 
to be $\sim 0.1$, indicating good convergence. Third order contributions
to the free energy are typically a few meV/atom, and comparable to their
statistical errors. The statistical error 
in the second order free energy of Al is 0.4 meV/atom with 
only 100 evaluations of the
{\it ab initio} energy. These results suggest that the perturbation
series allows for efficient and accurate evaluation of the free energy 
for condensed phases.

\end{abstract}

\maketitle

\section{Introduction}

Standard methods for determining the free energy for a given potential energy 
surface include thermodynamic and coupling constant 
integration \cite{allen_tildesley,understanding_molecular},  umbrella sampling
\cite{torrie_umbrella_1977} and related generalized ensemble 
methods \cite{berg_multicanonical_1992,wang_landau_2001}. The present work
is motivated by the pursuit of efficient methods for determining bulk free
energies of substances from {\it ab initio} potential energy surfaces.
Classical statistics will be considered here, as is appropriate for liquids
and anharmonic solids, except for the lightest elements.
Standard methods may not be well suited to this application because they require
a large number of evaluations of the expensive {\it ab initio} energy.
The perturbative expansion due to Zwanzig~\cite{zwanzig_pert} expresses the
Helmholtz  free energy $F$ of the system of interest, 
with potential energy function
$U$,  with respect to that
of a known reference system with potential energy $U_0$,
\begin{equation}
\Delta F = F - F_0  = \sum_{n=1}^{\infty} \frac{(-\beta)^{n-1}}{n!} \kappa_n 
\label{pert_exp}
\end{equation}
where $\beta$ is the inverse temperature and the $\kappa_n$ are cumulants of the 
potential energy difference $\Delta U = U-U_0$. Units where Boltzmann's constant
$k_B = 1$ will be used in the formulae that follow. Results and figures 
will use conventional temperature units. The first few of these cumulants are 
\begin{eqnarray}
\kappa_1 & = & \left\langle \Delta U \right\rangle_0 \nonumber \\
\kappa_2 & = & \left\langle \Delta U^2 \right\rangle_0
                - \left\langle \Delta U \right\rangle_0^2 \nonumber \\
\kappa_3 & = & \left\langle \Delta U^3\right\rangle_0 
    - 3 \left\langle \Delta U \right\rangle_0 
                 \left\langle \Delta U^2 \right\rangle_0
          +2 \left\langle \Delta U \right\rangle_0^3  \, ,
\label{cumulants}
\end{eqnarray}
where the notation $\langle \rangle_0$ denotes an expectation value 
in the canonical ensemble of the reference system. 
This expansion has some advantageous properties for the present application.
Ensemble averages in the reference canonical ensemble can be computed by
Monte Carlo sampling of the reference potential, which may be a simple pair
potential, or similar model, and thus be orders of magnitude faster to
evaluate than the {\it ab initio} potential. Equilibration and the generation
of statistically independent configurations is carried out on the fast reference
potential, and {\it ab initio} energies evaluated only for a greatly reduced
set of configurations.
Further gains in statistical efficiency result from the correlated fluctuations
of the reference and system potentials, which reduce the statistical
errors in the dominant first order  term.

Some earlier findings indicating the promise of the approach include an 
investigation of liquid Cu represented by an embedded atom model (EAM) potential
\cite{greeff_mcpert}, with a $1/r^{12}$ potential as the reference. 
There it was found that the series truncated at second order
has errors $< 1$~meV/atom without extensive fine-tuning of the reference
potential. Statistical errors of a few meV/atom were obtained with 100 samples.
In another work, {\it ab initio} energies were used for liquid Mg 
at ambient melting conditions \cite{greeff_liq_metal}.
The excess entropy obtained at second order was -3.18(15) $k_B$/atom 
(the number in parenthesis denotes the one standard deviation statistical 
error in the final digits),
compared to an experimental value of -3.37 $k_B$/atom.

Due to the factors $\beta^{n-1}$ in the series Eq.~(\ref{pert_exp}),
the expansion is usually called a high-temperature 
series \cite{zwanzig_pert,zhou_solana,dubinin_2014}. Zwanzig included this 
phrase in the title of his paper~\cite{zwanzig_pert}. It might be expected that
at low temperatures, truncating the series would lead to large errors,
and its application to condensed phases would be suspect. 

In the case of a hard sphere reference potential, the cumulants are 
temperature-independent and the series, Eq. (\ref{pert_exp}) is indeed
a high-temperature expansion. While the hard sphere reference system has 
played a historically important role in the theory of 
liquids \cite{hansen_mcdonald,barker_henderson,zhou_solana},
it has no particular advantage for the applications of interest here.
For more realistic, smooth reference potentials, the $\kappa_n$ will
depend on $T$, and this dependence, together with the $\beta^{n-1}$
factors will determine the temperature behavior of the series. To the 
extent that $\kappa_{n+1}/\kappa_n \propto T$, the convergence of the series
is independent of $T$. Dimensional analysis suggests $\kappa_n \propto T^n$.

The rest of this paper explores role of temperature in the convergence
of the free energy perturbation series, Eq.~(\ref{pert_exp}) for condensed
phases with smooth reference potentials. The first case considered is
where both the system and reference potentials are harmonic. In this case,
a closed form expression can be derived for $\kappa_n$, and it is found
that $\kappa_n \propto T^n$ for $n \ge 2$ and the convergence
of the series is independent of $T$. Truncation of the series results in
a temperature-independent error in the entropy. Subsequently, two more 
realistic cases are considered. The $\kappa_n$ for $n \le 3$ are evaluated
numerically by Monte  Carlo sampling of the reference canonical distribution.
The first is a model of liquid Cu, represented by an EAM 
potential \cite{voter_chen}, with a $1/r^{12}$ pair potential as the 
reference system. It is found that with fixed reference potential strength,
the convergence of the series gets {\it worse} with increasing temperature,
while with the reference potential optimized at each temperature, the convergence
is nearly independent of $T$. The final example is that of fcc Al
with DFT energies and a harmonic reference potential. In this case, the ratio
of the third order to the second order term in the free energy expansion
is found to vary from 0.13 (39) at 300~K to 0.04 (28) at 900~K, again
suggesting good convergence at low temperature.

\section{Cumulant Expansion}

Some further properties of the expansion, Eq. (\ref{pert_exp}) will be 
used in the following. The first is that under a constant 
shift $\Delta U \rightarrow \Delta U+c$,
$\kappa_1 \rightarrow \kappa_1+c$ while all $\kappa_n$ for $n>1$ are
unchanged. In applications, there are often large offsets between the system and
reference potentials due to different zero energy states. This property means
that these offsets are accounted for by the first order term in the series. 
The second property of interest is the variational 
property \cite{hansen_mcdonald,barker_henderson}
of the first order approximation
\begin{equation}
F_1 = F_0 + \left\langle \Delta U \right\rangle_0 \ge F
\label{variational}
\end{equation}
which allows optimization of the reference potential by minimizing $F_1$.

\section{Harmonic Case}

Consider the case where both the system and the reference are harmonic
\begin{eqnarray}
U & = & \phi + \frac{1}{2}  \mathbf{u}^T \mathbf{M u}   \nonumber \\
U_0  & = & \phi_0 + \frac{1}{2} \mathbf{u}^T \mathbf{M_0 u}  
\label{uharmonic}
\end{eqnarray}
where $\mathbf{u}$ is a $3N$-dimensional displacement vector from the 
local potential minimum, whose energy
is $\phi$. The matrices $\mathbf{M}$ and $\mathbf{M_0}$ are symmetric and
positive definite.
They are not assumed to commute.
This model is of interest as a realistic approximation for solids. 

There are a number of ways to show that for this harmonic case, the cumulants
$\kappa_n \propto T^n$ to all orders. 
It turns out to be possible to work out  a closed form expression for the
full expansion. Starting from the free energy difference in terms of the 
partition function, we have,
\begin{eqnarray}
\Delta F 
  & = &  -T \ln \frac{ \int d\mathbf{u} \exp{\left[-\beta 
       \left( \phi + \frac{1}{2} \mathbf{u}^T \mathbf{M u} \right)\right]} }
   { \int d\mathbf{u} \exp{\left[-\beta \left( \phi_0 + \frac{1}{2} 
                                \mathbf{u}^T \mathbf{M_0 u} \right)\right]} } \, .
\\
& & \nonumber 
\end{eqnarray}
Using a standard result for Gaussian integrals \cite{zinn-justin_path}, this is
\begin{eqnarray}
\Delta F & = & \phi - \phi_0 + 
              \frac{T}{2}\ln\frac{\rm{Det}\left( \mathbf{M}\right)}
                                        {\rm{Det}\left( \mathbf{M_0}\right)}
\nonumber \\
  & = & \phi - \phi_0 \nonumber \\
  & &  + \frac{T}{2}\ln \rm{Det}\left( 1 + \mathbf{M_0}^{-1/2}
                    \, \Delta\mathbf{M} \, \mathbf{M_0}^{-1/2} \right)
\end{eqnarray}
where $\Delta\mathbf{M} = \mathbf{M} - \mathbf{M_0}$. Then we can use the 
identity $\ln \rm{Det} \left( \mathbf{A} \right) 
= \rm{Tr} \ln\left( \mathbf{A} \right)$ to write
\begin{eqnarray}
\Delta F & = & \phi - \phi_0 + \frac{T}{2} \rm{Tr} 
\ln \left( 1 + \mathbf{M_0}^{-1/2} \, \Delta\mathbf{M} \, \mathbf{M_0}^{-1/2} \right) \, .
\end{eqnarray}
Finally, expanding the logarithm, we have
\begin{eqnarray}
\Delta F & = & \phi - \phi_0 \nonumber \\
     & & +  \frac{T}{2} \sum_{n=1}^\infty \frac{(-1)^{n-1}}{n}
     \rm{Tr} \left[ \left(\mathbf{M_0}^{-1/2} \, \Delta\mathbf{M}  \,
                                            \mathbf{M_0}^{-1/2} \right)^n \right]
                     \nonumber \\
         & = & \phi - \phi_0 +  \frac{T}{2} \sum_{n=1}^\infty \frac{(-1)^{n-1}}{n}
     \rm{Tr} \left[ \left(\mathbf{M_0}^{-1} \Delta\mathbf{M} \right)^n \right]
\label{harmonic_exp}
\end{eqnarray}
Eq. (\ref{harmonic_exp}) is an expansion of $\Delta F$ in powers of 
$\Delta U \propto \Delta \mathbf{M}$, and is therefore
equivalent to Eq. (\ref{pert_exp}) for the harmonic  case. It is straightforward
to check that a direct evaluation of the first few orders with the formulae
Eqs. (\ref{cumulants}) yields the same results as Eq. (\ref{harmonic_exp}).
So, by comparison of Eq. (\ref{harmonic_exp}) to Eq. (\ref{pert_exp}), we
can read off, 
\begin{eqnarray}
\kappa_1 & = & \phi - \phi_0 + \frac{T}{2} 
        \rm{Tr} \left( \mathbf{M_0}^{-1} \Delta\mathbf{M} \right)
       \nonumber \\
\kappa_n & = & T^n \frac{(n-1)! }{2}  
    \rm{Tr} \left[ \left(\mathbf{M_0}^{-1} \Delta\mathbf{M} \right)^n \right],
       \, n \ge 2
\\ \nonumber 
\end{eqnarray}
The first cumulant incorporates the constant shift $\phi - \phi_0$ as well
as a term $\propto T$, while the higher cumulants have $\kappa_n \propto T^n$.
The temperature factors out of the series in Eq. (\ref{harmonic_exp})
so $T$ does not enter into its convergence  
at all. Convergence of  the series requires that
the eigenvalues $\lambda_i$ of the matrix $\mathbf{M_0}^{-1} \Delta\mathbf{M}$
satisfy $-1 < \lambda_i \le 1$, and is thus  
contingent only on properties of the reference and system potential surfaces. 
Truncation of the perturbation series results in a temperature-independent
error in the entropy.

\section{Liquid Copper}

The case of liquid copper,
represented by an embedded atom potential \cite{voter_chen}, with a $1/r^{12}$
reference potential, is considered as  a realistic example of a dense fluid. 
The first three cumulants are evaluated by direct Monte Carlo sampling 
of the canonical distribution of
the reference system. 
The system consists
of 256 atoms in a cubic periodic cell. The reference is potential is
\begin{equation}
U_0 = \sum_{i<j} \epsilon \left( \frac{\sigma}{r_{ij}} \right)^{12} \, .
\end{equation} 
The parameter $\sigma $ is taken to be 0.14 nm. This is arbitrary, since
only the combination $\epsilon \sigma^{12}$ enters, but it is convenient,
and allows $\epsilon$ to take on reasonable values. The parameter $\epsilon$ 
is treated as variational.
As described above, the cumulants $\kappa_n$ for $n \ge 2$ are indicative of 
the convergence of the series, while the first cumulant contains a large 
contribution from the cohesive energy, analogous to the term $\phi-\phi_0$
in the harmonic case. To the extent that $\kappa_2/T^2$ $\kappa_3/T^3$
are independent of $T$, this case is similar to the harmonic case, suggesting
that convergence is independent of $T$.
Higher order cumulants
have increasing statistical errors, so third order was chosen as a 
practical limit.

\begin{figure}
\noindent
\includegraphics[scale=0.33]{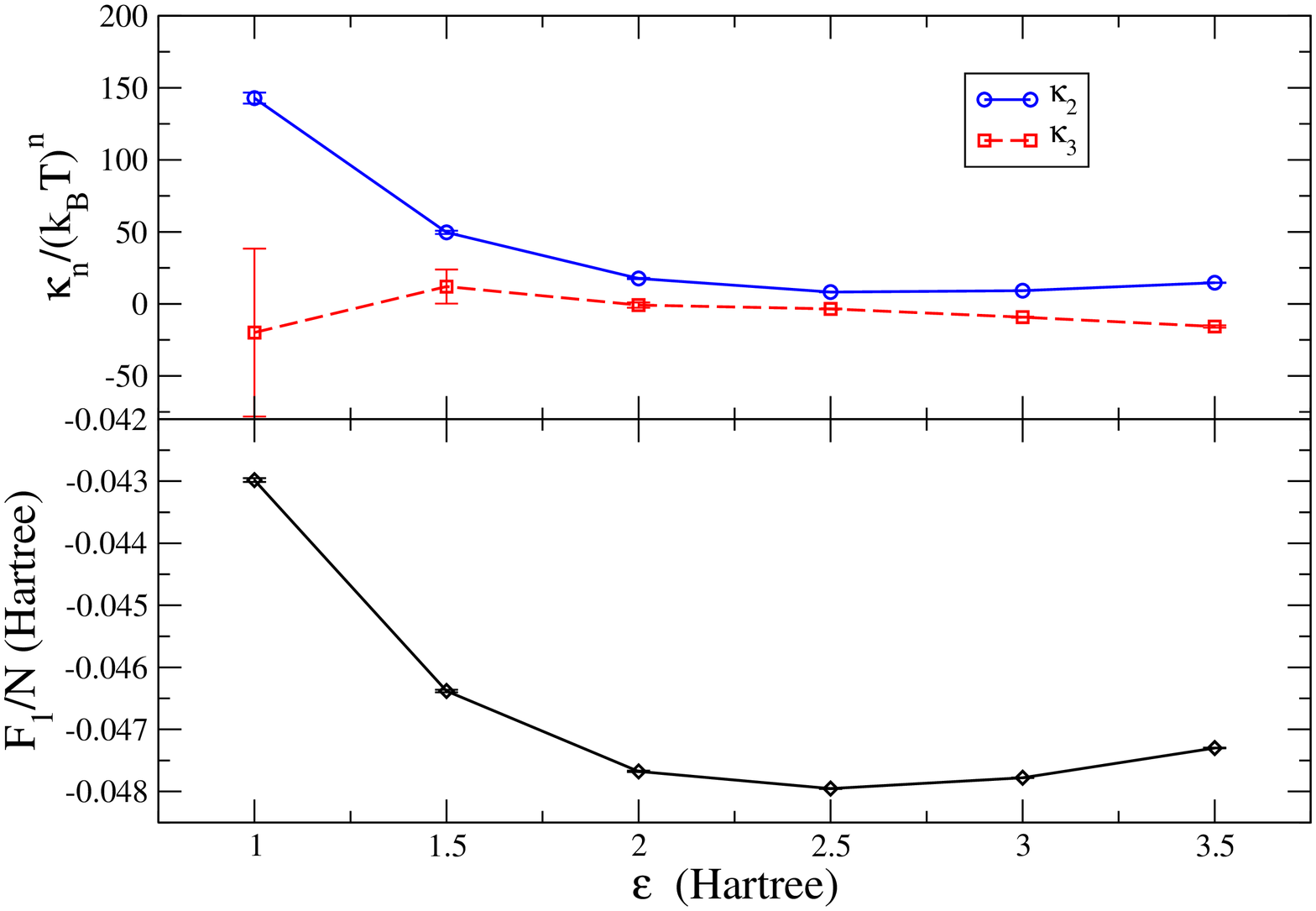}
\caption{Dependence of the cumulants $\kappa_3$ and $\kappa_3$
on the reference potential parameter $\epsilon$. The system is liquid Cu at $T = 7200$ K and $\rho = 0.08467$ \AA$^{-3}$. The lower panel shows the
first order excess free energy $F_1$, which obeys the variational inequality.
The variational optimum is $\epsilon \approx 2.5$ $E_h$, which also 
corresponds to the minimum of $\kappa_2$.}
\label{f1_opt_kn}
\end{figure}

Two cases are considered, one with fixed $\epsilon$, and one with
$\epsilon$ approximately optimized at each temperature.
The optimization process was to vary $\epsilon$ so as to minimize 
$\kappa_2$.
Figure \ref{f1_opt_kn} illustrates the optimization process
for liquid
Cu at $T = 7200$ K and $\rho = 0.08467$ \AA$^{-3}$.  
The free energy of the reference system as parameterized by
Young and Rogers \cite{young_r12} was used. At this temperature, the optimum
value of $\epsilon$, in the sense of minimizing $F_1$, is approximately 
2.5 $E_h$, where $E_h$ denotes the Hartree energy, $E_h = 27.2114$~eV. 
At this value, $\kappa_2$ is also minimized, and $\kappa_3$
is small. So, variational optimization corresponds approximately to
optimizing the convergence of the series. Subsequently, $\epsilon$
is optimized by minimizing $\kappa_2$.

\begin{figure}
\noindent
\includegraphics[scale=0.33]{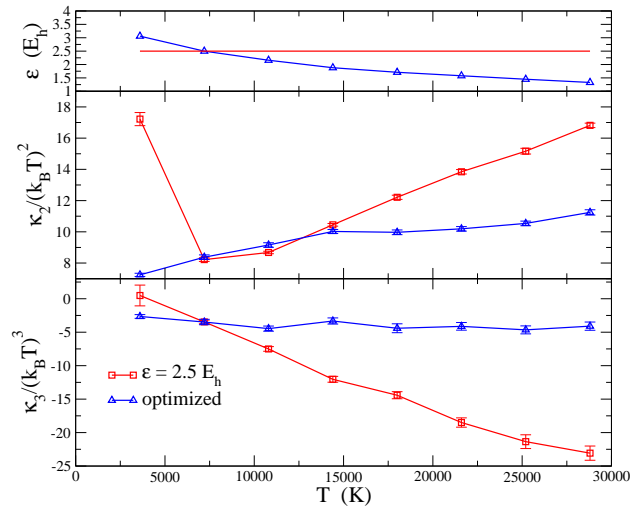}
\caption{Variation of the cumulants $\kappa_2/T^2$ and $\kappa_3/t^3$
with $T$ for EAM liquid Cu at $\rho = 0.08467$ \AA$^{-3}$. 
Red curves are for the reference potential parameter $\epsilon$ 
fixed at 2.5 $E_h$, and red curves are for $\epsilon$ optimized at each $T$.}
\label{kn_of_t}
\end{figure}

Figure \ref{kn_of_t} shows the second and third cumulants as functions of
temperature, evaluated with both fixed and optimized $\epsilon$. 
The plots show the dimensionless cumulants, $\kappa_n/T^n$, as 
appear in the expansion for $\beta \Delta F$. 
With fixed parameters, the dimensionless second cumulant 
varies non-monotonically with 
$T$, initially decreasing, then increasing. The dimensionless third cumulant 
increases in
magnitude with $T$ over the whole range. With optimized $\epsilon$, the variation
of the cumulants is greatly reduced, with $\kappa_2/T^2$ increasing
slowly with $T$ and $\kappa_3/T^3$ remaining nearly constant. This indicates that
the convergence of the series is nearly independent of $T$. 
 The absolute ratio of the third to second order terms,
$1/3 \left| \beta^3 \kappa_3/\beta^2 \kappa_2 \right|$ ranges from 0.01(27) to 0.46(19)
from low to high temperature, with fixed $\epsilon$, and from 0.12(12) to
0.12(16) with optimized $\epsilon$.
At the lowest temperature, 3600~K, the third order contribution to the free
energy is $2.0 \times 10^{-5}$~$E_h$/atom (0.5 meV/atom), indicating that
it is very small in practical terms.

\section{Al Crystal}

The final example considered here is Al in the fcc crystalline state. 
The system potential energy surface is defined by {\it ab initio}
DFT energies, and the reference potential is harmonic. 
The DFT potential energy surface is evaluated using the plane wave 
code Abinit \cite{abinit1,abinit2} for a 108 atom cell consisting of $3^3$ 
conventional fcc cells. The JTH PAW data \cite{PAW_JTHv1.0} was used,
and ecut was set to 20 $E_h$. Occupations were determined 
using the ``cold smearing'' option (occopt = 4) with the smearing 
parameter set to 0.010 $E_h$. The Brillouin zone was sampled with a $2^3$
shifted $k$-point mesh. The tolerance for SCF convergence was $10^{-6} E_h$.

The reference potential was harmonic, with force constants defined by an 
axially symmetric model \cite{gilat_al_axial_1966}. 
Ref. \cite{gilat_al_axial_1966} fitted an eighth 
neighbor model to the phonon dispersion data. In order to test the perturbative 
expansion, the force constants were truncated at second neighbors. The
perturbation series then has to account for deficiencies in the short-range 
model, as well as anharmonic effects.
The reference free energy is 
$F_0 = 3 N T \left[ \ln\left(\hbar \omega_0/T \right) -1/3 \right]$, 
where  
$\omega_0 = \exp \left[ \langle \ln \omega \rangle_{\rm BZ} + 1/3 \right]$
is the logarithmic moment of the
phonon frequencies and $\langle \rangle_{\rm BZ}$ denotes an average 
over wave vectors and polarizations in the Brioullin zone.
This average is evaluated on an $8^3$ shifted $k$-point grid, which gives
convergence to 1 part in $10^4$.

Nuclear coordinates were sampled from the canonical distribution of the
reference harmonic model using standard Metropolis 
Monte Carlo \cite{allen_tildesley}.
The configurations were equilibrated over 2000 sweeps consisting of one
attempted move of each atom. Then, configurations were sampled every 500 sweeps
for evaluation of their DFT energy. In the present case, 100 samples were 
generated for each thermodynamic state. These configurations were distributed 
to separate directories for DFT energy calculations. The results were gathered 
afterwords, and the terms in the perturbation series were evaluated  
as ensemble averages over the sample. Statistical errors were estimated
by breaking up the sample into four sub-samples. Reported error bars are 
one standard deviation. This example is meant to be representative  
of what would be feasible for a broad scan of thermodynamic states using
an {\it ab initio} potential energy surface.

\begin{figure}[t]
\noindent
\includegraphics[scale=0.33]{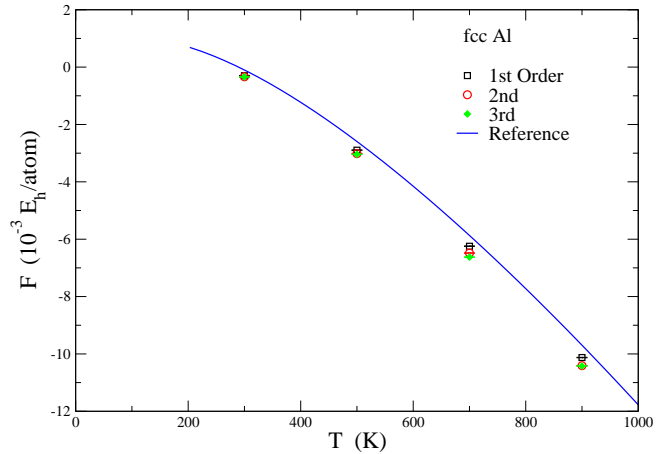}
\caption{Free energy of fcc Al. Solid blue curve: harmonic reference. Black squares: 1st order, red circles: second order, green diamonds: 3rd order perturbation
theory. The zero of energy is the energy of the fcc crystal.}
\label{al_foft}
\end{figure}

The resulting Helmholtz free energies for Al at the fixed lattice
parameter $a = 7.646 a_0$ (4.046 \AA)  are shown in figure \ref{al_foft}.  
The second order contribution to the free energy increases 
in magnitude from $0.4-2.8\times10^{-4}$
$E_h$/atom (1-7.6 meV/atom) over the temperature range from 300 to 900~K.
The third order correction to the free is generally small, around 0.1 $E_h$/atom
(3 meV/atom)  or less, and comparable 
to its statistical uncertainty. At 300~K, the ratio of the third order to
the second order term is 0.13 (39), while at 900 K the ratio is 0.04 (28)
indicating good convergence. At the 900 K, statistical errors in the 
second order free 
energy are   $1.6 \times10^{-5}$~$E_h$/atom (0.4 meV/atom).
All indications are that the truncation error of the the perturbation 
series and the statistical errors are small, on the order of a few meV/atom.

\section{Conclusions}

The perturbation series introduced by Zwanzig \cite{zwanzig_pert}
has properties which are advantageous for the efficient evaluation 
of bulk free energies from {\it ab initio} potential energy surfaces.
However, the presence of increasing powers of the inverse temperature
suggests that the series might not be expected to converge well for 
condensed phases. It has been shown here, for the case that both the
system and reference potential are harmonic, that the cumulants
$\kappa_n \propto T^n$ for $n \ge 2$, and thus that the convergence
of the series is independent of $T$. In more realistic examples
of liquid Cu and fcc Al, using smooth reference potentials, 
numerical results indicate that the truncation error at second order
is a few meV/atom or less. 

There is a trade-off between the accuracy of a high order expansion 
and the statistical
error, which increases for high order cumulants. The results here suggest that
a practical procedure is to use the second order approximation, while
evaluating the third order term to monitor convergence. In the Al results
obtained here, the third order term was within one standard deviation of
zero in most cases.

In the case of fcc Al,
the  statistical errors of the second
order free energy were less than 1 meV/atom with only 100 evaluations 
of the {\it ab initio} energy. This is notable 
because to achieve this statistical accuracy in the energy, for example,
with {\it ab initio} molecular dynamics would usually require a much
larger number of {\it ab initio} force and energy evaluations.
This  efficiency stems from two main sources. Because the ensemble averages
being evaluated are over the canonical distribution of the {\it reference}
system, equilibration and the generation of independent samples is done
on the fast reference potential. Because the energy fluctuations of the
reference and system potentials are correlated, the fluctuations of the 
energy difference are much smaller than the fluctuations of the 
energies themselves. In summary, the cumulant expansion of the free energy,
when used with realistic, smooth, reference potentials, 
affords an efficient method with small errors for low-temperature states 
such as solids and dense liquids.

\bibliography{lowt.bib}

\end{document}